\documentclass[a4paper, amsfonts, amssymb, amsmath, reprint, showkeys, twoside]{revtex4-1}
\usepackage[english]{babel}
\usepackage[utf8]{inputenc}
\usepackage{color}
\usepackage[colorinlistoftodos, color=green!40, prependcaption]{todonotes}
\usepackage{amsthm}
\usepackage{mathtools}
\usepackage{physics}
\usepackage{xcolor}
\usepackage{graphicx}
\usepackage[left=23mm,right=13mm,top=35mm,columnsep=15pt]{geometry} 
\usepackage{adjustbox}
\usepackage{placeins}
\usepackage[T1]{fontenc}
\usepackage{lipsum}
\usepackage{csquotes}\usepackage[pdftex, pdftitle={Article}, pdfauthor={Author}]{hyperref} % For hyperlinks in the PDF

\begin{document}
	\title{{Synchronization dynamics in non-normal networks: the trade-off for optimality}}
	
		\author{Riccardo Muolo$^\dagger$}
%	\email[Correspondence email address: ]{riccardo.muolo@unamur.be}% Your name

		\author{Timoteo Carletti$^\dagger$}
%	\email[Correspondence email address: ]{timoteo.carletti@unamur.be}% Your name

		\author{James P. Gleeson$^\ddagger$}
	%\email[Correspondence email address: ]{malbor.asllani@ul.ie}% Your name

	\author{Malbor Asllani$^{\dagger, \ddagger}$}
	%\email[]{malbor.asllani@ul.ie}% Your name
	
		\affiliation{$\dagger$ Department of Mathematics and naXys, Namur Institute for Complex Systems, University of Namur, rempart de la Vierge 8, B 5000 Namur, Belgium }
	
	\affiliation{$\ddagger$ MACSI, Department of Mathematics and Statistics, University of Limerick, Limerick V94 T9PX, Ireland }
	
	\date{\today} % Leave empty to omit a date
	
\begin{abstract}
Synchronization is an important behavior that characterizes many
natural and human made systems composed by several interacting units. It can be found in a broad spectrum of applications, ranging from neuroscience to power-grids, to mention a few.
Such systems synchronize because of the complex set of coupling they exhibit, the latter being modeled
by complex networks. The dynamical behavior of the system and the topology of the underlying network are strongly intertwined, raising the question {of} the optimal architecture that makes synchronization robust. The Master Stability Function (MSF) {has} been proposed {and extensively studied} as a generic framework to tackle synchronization problems. Using this method{, it} has {been} shown that for a class of models, synchronization in strongly directed networks is robust to external perturbations. In this paper, our approach is to transform the non-autonomous system of coupled oscillators into an autonomous one, showing that previous results are model-independent. Recent findings indicate that many real-world networks are strongly directed, being potential candidates for optimal synchronization. {Inspired by the fact that highly directed networks are also strongly non-normal,} in this work, we address the matter of non-normality by pointing out that standard techniques, such as the MSF, may fail in predicting the stability of synchronized behavior. These results lead to a trade-off between non-normality and directedness that should be 
 properly considered when designing an optimal network, enhancing the robustness of synchronization.
\end{abstract}	
	%\keywords{synchronization dynamics, non-normal networks, optimal networks, master stability function}

	\maketitle
	
\section{Introduction}
\label{sec:intro}
\noindent
Systems in nature are often constituted by a large number of small parts that continuously interact with each other~\cite{newman, arenas_rev}. Although it might be possible to accurately know the dynamics that characterize each of the individual constituents, it is, in general, nontrivial to figure out the collective behavior of the systems as a whole {resulting from the individual/local interactions}. A relevant example is {provided by a system composed by}  an ensemble of {coupled non-linear} oscillators, {that behave at unison driven by the non-local interaction, {then the system is said to be} synchronized~\cite{strogatz, arenas_rev}. {Synchronization has been {extensively} studied in network science as a {paradigm} of dynamical processes on a complex network, mainly due to the essential role of the coupling topology in the collective dynamics}~\cite{newman}.
{Its generic formulation allowed researchers to use it to model} several applications, ranging from {biology, e.g., neurons firing in synchrony,} to {engineering, e.g.,} power grids~\cite{pik}. {The ubiquity of synchronization} in many natural or artificial systems {has naturally raised questions about} the stability and robustness of synchronized states~\cite{skardal1, skardal2, mott_nish1, mott_nish2}.
In their seminal work, Pecora \& Caroll~\cite{pecora} introduced a method known as Master Stability Function (MSF) {to help} understand the role that the topology of interactions has on {system} stability. {Assuming a diffusive-like coupling among the oscillators, the MSF relate{s} the stability of the synchronous state to the nontrivial spectrum of the (network) Laplace matrix;} in particular, {it has been proven that the latter} should lie in the region where the Lyapunov exponent that characterizes the MSF takes negative values~\cite{arenas_rev, bara}. {For a family of models (e.g., R\"ossler, Lorenz, etc.) whose stable part of the MSF has a continuous interval where the (real part of the) Laplacian eigenvalues can lie, it has been proven that they maximize their stability once the coupling network satisfies particular structural properties.} Such optimal networks should be directed spanning trees and without loops~\cite{mott_nish1, mott_nish2}. These networks have the peculiarity of {possessing} {a degenerate} spectrum of the Laplacian matrix and laying in the stability domain {provided by} the Master Stability Function. The Laplacian degeneracy {is also often associated with a real spectrum} or with considerably low imaginary parts compared to the real ones \cite{degen, trefethen}.

The vast interest in complex networks in recent years has also provided an abundance of data {on} empirical networked systems that initiated a large study of their structural properties~\cite{newman}. From this perspective, it has been recently shown that many real networks are strongly directed, {namely} {they possess} a high asymmetry adjacency matrix~\cite{malbor_teo}. Most of these networks present a highly hierarchical, almost-DAG (Directed Asymmetric Graph) structure. This property potentially makes the real networks suitable candidates for {optimally synchronized dynamical systems defined on top of them}. Another aspect which is unavoidably associated with the high asymmetry of real networks, is their non-normality~\cite{malbor_teo}, namely {their adjacency} matrix $\mathbf{A}$ satisfies the condition $\mathbf{A}\mathbf{A}^T\neq \mathbf{A}^T\mathbf{A}$~\cite{trefethen}. The non-normality can be critical for the dynamics of networked systems~\cite{top_resilience, malbor_teo, me, duccio_NN, duccio_stoch, baggio}. In fact, in the non-normal dynamics regime {a finite perturbation about a stable state} can undergo a transient instability~\cite{trefethen} which {because of the non-linearities could never be reabsorbed}~\cite{top_resilience, malbor_teo}. The effect of non-normality in dynamical systems has been studied in several contexts, such as hydrodynamics~\cite{trefethen2}, ecosystems stability~\cite{neub_cas}, pattern formation~\cite{murray}, chemical reactions~\cite{mott_nn}, etc. {However, it is only recently that} the ubiquity of non-normal networks and the related dynamics have been put to the fore~\cite{top_resilience, malbor_teo, me, duccio_NN, duccio_stoch,baggio}. In this paper, we will elaborate on these lines showing the {impact of non-normality on the stability of a synchronous state}. We first show that a strongly non-normal network has, in general, a spectrum {very close to a real one} and that this in principle should imply a {larger domain of parameters for which stability occurs, for systems with a generic shaped MSF}. For illustration purposes, we will consider the Brusselator model \cite{prigo, galla}, a two-species system with {a} {discontinuos interval of stability in the MSF representation.} {We will also {examine the limiting cases of} our analysis {to two simple} network models~\cite{malbor2}, namely a (normal) bidirected} circulant network {and a (non-normal) chain}, both with tunable edge weights {in such a way to allow a continuous adjustment respectively of the directedness and non-normality}.

{The MSF relies on the computation of the (real part of the maximum) Lyapunov exponent, and thus in the case of time-dependent systems, it does not possess the full predictability power it has in the autonomous case {(fixed point in/stability)}. For this reason, we will use a homogenization method, whose validity is limited to a specific region of the model parameters, allowing {us} to transform the linearized periodic case problem into a time-independent one \cite{averaging}. This way, we remap our problem to an identical one studied in the context of pattern formation in directed networks where spectral techniques provide {significant} insight~\cite{malbor,malbor2}.}
Such an approach allows {us} on one side {to assess} the quantitative evaluation of the role of the imaginary part of the Laplacian spectrum in the {stability problem}. On the other {it} permits the use of numerical methods, such as the pseudo-spectrum~\cite{trefethen} in the study of the non-normal dynamics. To the best of our knowledge, {this is the first attempt to use such techniques in the framework of time-varying systems, being} the theory of non-normal dynamical systems limited so far to autonomous systems~\cite{trefethen}. As expected, the non-normality plays against the stability of the synchronized ensemble of oscillators. Furthermore, a high non-normality translates to a high spectral degeneracy, which brings to a large pseudo-spectrum, indicating a high sensibility towards the instability.

{Clearly, the directionality and the non-normality stand on two parallel tracks regarding the stability of synchronized states and their robustness. As a conclusion of our work, we show that the most optimal design should be looked at as a trade-off between a high and low directionality/non-normality. Such choice should depend either on the magnitude of perturbation or the ratio directed vs. non-normal of the network structure.}

\section{Optimal synchronization: Directed vs. Non-normal networks}
\noindent
{We consider a network {constituted of} $N$ nodes (e.g., the idealized representation of a cell), and we assume a metapopulation framework, where the species dynamics inside each node is described by the \textit{Brusselator} model, a portmanteau term for Brussels and oscillator. It has been initially introduced by Prigogine \& Nicolis to capture the autocatalytic oscillation~\cite{prigo} phenomenon, resulting from a Hopf bifurcation curve in the parameter plane.
 %occurring at $c=b-1$~\cite{prigo, galla}. 
 This will be the framework we will consider in the following, neglecting thus the fixed point regime. Specie can migrate across nodes with a diffusion-like mechanism}. In formulae, this model translates to a reaction-diffusion set of equations:%\vspace*{-.25cm}
\begin{equation}
\begin{cases}
\dfrac{d\varphi_i}{dt}=1-(b+1)\varphi_i+c\varphi_i^2\psi_i+D_\varphi\displaystyle \sum_{j=1}^N \mathcal{L}_{ij}\varphi_j\\
\dfrac{d\psi_i}{dt}=b\varphi_i-c\varphi_i^2\psi_i+D_\psi \displaystyle \sum_{j=1}^N \mathcal{L}_{ij}\psi_j\,, \; \forall i=1,\dots ,N,
\end{cases}
\label{eq:RD}
\end{equation}
where $\varphi_i$ and $\psi_i$ indicate the concentration of the two species per node, $D_\varphi$, $D_\psi$ are their corresponding diffusion coefficients, and $b$, $c$ are the model parameters. {The coupling is represented by the matrix $\boldsymbol{\mathcal{W}}$, whose {non-negative} entries $\mathcal{W}_{ij}$ represent the strength of the edge pointing form node $j$ to node $i$.} The entries of the Laplacian matrix $\boldsymbol{\mathcal{L}}$ are given by $\mathcal{L}_{ij}=\mathcal{W}_{ij}-k^{in}_i\delta_{ij}$ where $k_i^{in}=\sum_j \mathcal{W}_{ij}$ stands for the incoming degree of node $i$, i.e. the number of all the entering edges into node $i$. We want to emphasize here that many other coupling operators are also possible; nevertheless, most of them will reduce at the linear level to a Laplacian {involving the} differences of the observable {among} coupled nodes \cite{arenas_rev}, i.e., $\sum_{j=1}^N\mathcal{L}_{ij}x_j = \sum_{j=1}^N \mathcal{W}_{ij}(x_j-x_i)$. {This form ensures that the coupling is in action only when the observable assume different values {in two coupled nodes.}

\begin{figure*}[t!]
	\centering
	\includegraphics[width=\textwidth]{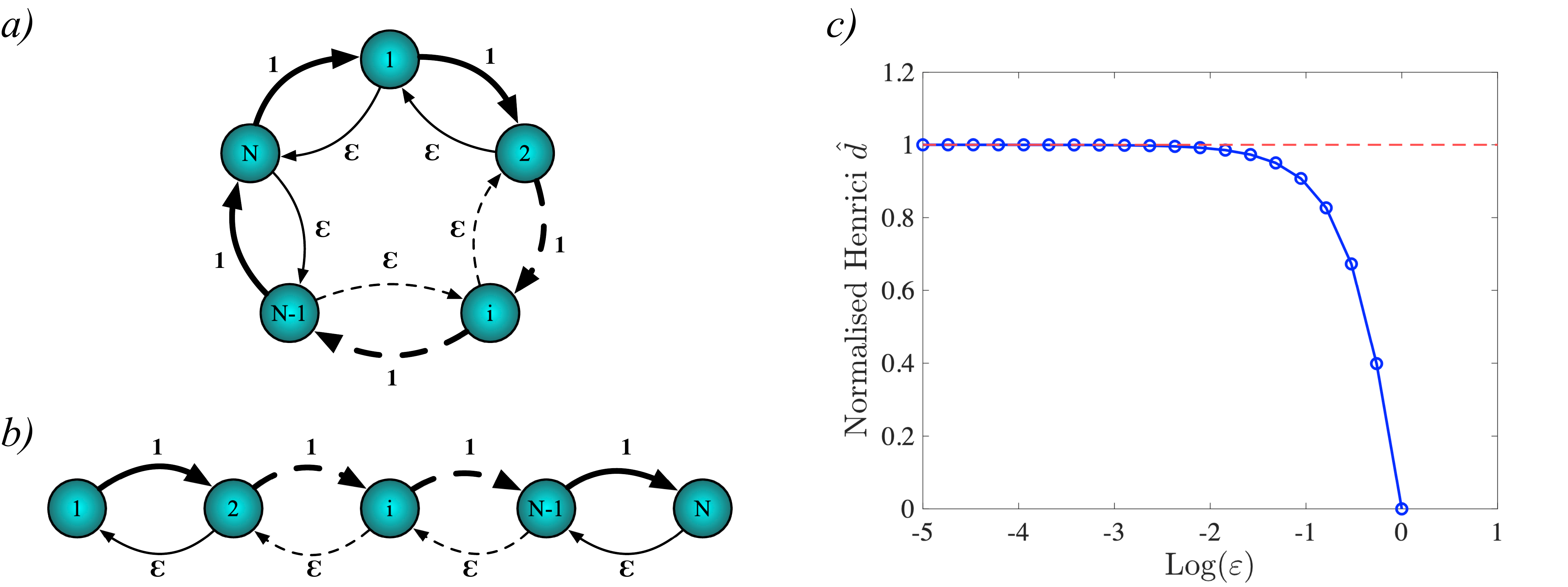}
	\caption{The network toy models for the case of a normal bidirectional circulant network{, panel $a)$}, and a non-normal bidirectional chain{, panel $b)$}. $c)$ Normalized Henrici's departure from non-normality as a function of tunning parameter $\epsilon$ {for the non-normal model}. We observe that starting from $0$, the network is symmetric, and the non-normality increases as the weight of the reciprocal edges decreases, taking the maximal value of non-normality in the limit when $\varepsilon=0$. In this case, the Laplacian spectrum is degenerate.}
	\label{fig:net_model}
\end{figure*}

The reason for choosing {such} a model, {as mentioned earlier,} is mainly {due to the discontinuous interval of the stability domain provided by the MSF of the problem} {(as it can be noticed in the inset of Fig. \ref{fig:confrontation} $a)$)}. To proceed with the stability analysis, we first need to {identify the homogeneous periodic solution, $\boldsymbol{\varphi}^*(t)$ and $\boldsymbol{\psi}^*(t)$, hereby called {the} {\em synchronized manifold} and then to} linearize the system around {this}. 
 {Let us introduce the} perturbations {for the $i$--th} node by $\delta\varphi_i$ and $\delta\psi_i$, {then the} linearized equations {describing} their evolution {are given by}:
\begin{eqnarray}
\dfrac{d(\delta\varphi_i)}{dt}&=&\left[f_{\varphi_i}\delta_{ij}+D_\varphi\sum_{j=1}^N \mathcal{L}_{ij}\right]\delta\varphi_j+f_{\psi_i}\delta\psi_i\nonumber\\
\dfrac{d(\delta\psi_i)}{dt}&=&g_{\varphi_i}\delta\varphi_i+\left[g_{\psi_i}\delta_{ij}+D_\psi  \sum_{j=1}^N \mathcal{L}_{ij}\right]\delta\psi_j\, \nonumber\\ \forall i=&1&,\dots ,N,\nonumber\\
\label{eq:lin_RD}
\end{eqnarray}
where the partial derivatives are given by $f_{\varphi_i}=-(b+1)+2c\varphi^*(t)\psi^*(t)$, $f_\psi=c\varphi^*(t)^2$, $g_{\varphi_i}=b-2c\varphi^*(t)\psi^*(t)$, and $g_{\psi_i}=c\varphi^*(t)^2$. Notice that the partial {derivatives} {of the reaction part} are evaluated {on the synchronized manifold}. This translates into a time-dependent Jacobian matrix due to the periodicity of the solutions {and thus {to a}} non-autonomous {linear system}. {To make a step forward let us introduce the following compact notation; let $\mathbf{x}=(\delta \varphi_1,\dots,\delta \varphi_N,\delta \psi_1,\dots,\delta \psi_N)^{\mathrm{T}}$ be the $2N$-dimensional perturbations vector, $\boldsymbol{\mathcal{D}}$ the diagonal diffusion coefficients matrix and $\boldsymbol{\mathcal{J}}(t)$ the time-dependent Jacobian matrix, hence Eq.~\eqref{eq:lin_RD} {can be rewritten as}
% rewrites as
\begin{equation}
\label{eq:compact}
\dot{\mathbf{x}}=\left(\boldsymbol{\mathcal{J}}(t)+\boldsymbol{\mathcal{D}}\odot\boldsymbol{\mathcal{L}}\right)\mathbf{x}\, ,
\end{equation}
where $\odot$ is the {coordinatewise} multiplication operator. Then we proceed by diagonalizing the linearized system using the basis of eigenvectors of the network Laplace operator $\boldsymbol{\mathcal{L}}$. Notice that this is not always possible because the Laplacian matrix of directed networks might not {have} linearly independent eigenvectors. We will assume such a basis to exist for the time being, and we will consider such an issue again {when} discussing the non-normal case. Denoting by $\boldsymbol{\xi}$ the transformed perturbations vector, Eq.~\eqref{eq:compact} {becomes}
\begin{equation}
\dot{\boldsymbol{\xi}}=\left(\boldsymbol{\mathcal{J}}(t)+\boldsymbol{\mathcal{D}}\odot\boldsymbol{\Lambda}\right)\boldsymbol{\xi},
\label{eq:MSF}
\end{equation}
where $\boldsymbol{\Lambda}$ denotes the diagonal matrix of the Laplacian eigenvalues. The (real part of the) largest Lyapunov exponent of Eq.~\eqref{eq:MSF}, known in the literature as the Master Stability Function~\cite{pecora, bara, arenas_rev, newman}, {is} thus a function of the eigenvalues $\boldsymbol{\Lambda}$. Let us stress that the study of the stability of a general non-autonomous system is normally not possible through the classical spectral analysis, and one has 
{therefore} to resort to the MSF.}

%\onecolumngrid
%\begin{widetext}

\begin{figure*}[t!]
	\centering
	\includegraphics[width=\textwidth]{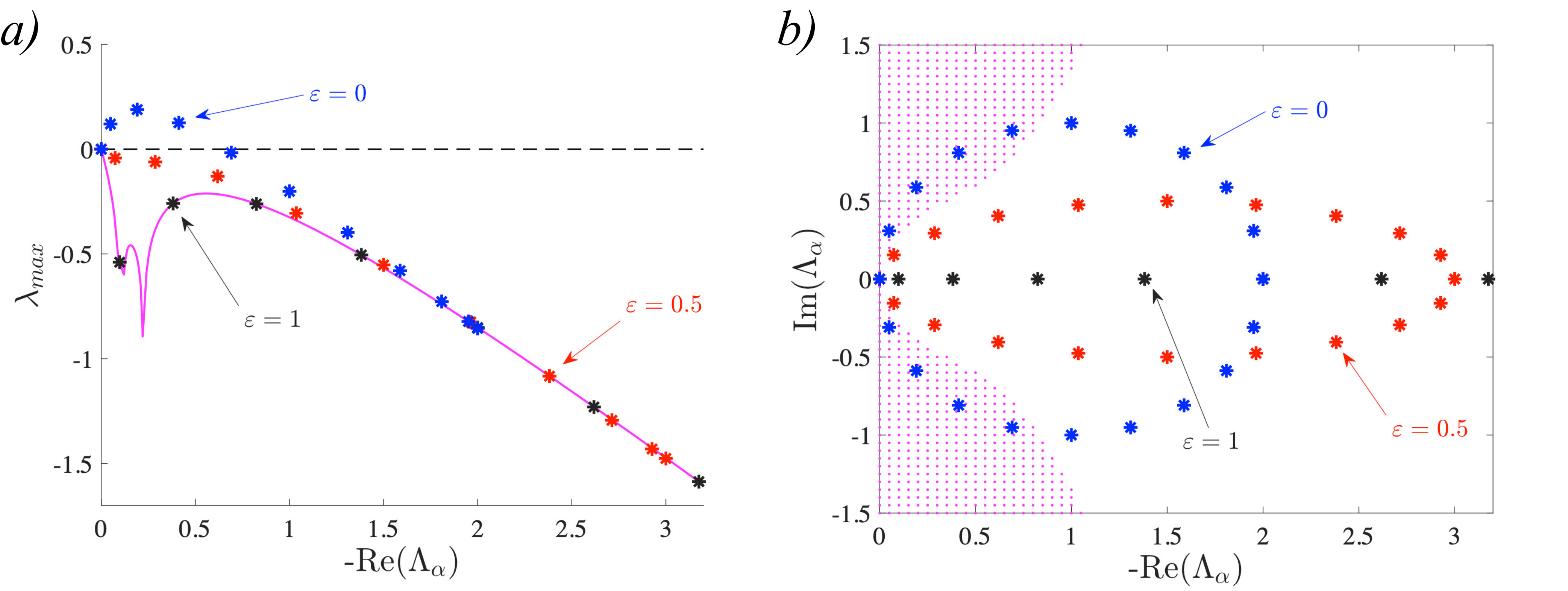}
	\caption{{$a)$} MSF for {the} Brusselator model with $b=2.5$, $c=1$ (limit cycle regime), $D_{\varphi}=0.7$, $D_{\psi}=5$ on a circulant network of $20$ node; {$\Lambda_\alpha$} indicates the {the Laplacian's eigenvalues}, of which we plot only the real part. {In this setting} the system should remain stable after a perturbation: in fact, when the network is symmetric ($\varepsilon=0$), the discrete MSF ({black} dots) lies on the continuous one ({magenta} line); however, when we introduce an asymmetry in the topology as $\varepsilon$ decreases ({red} and blue dots), the MSF reaches the instability region, and the system loses synchronization. {$b)$ The equivalent rapresentation in the complex domain where the instability region is shaded magenta and the discrete Laplacian spectrum is denoted by the symbols. For the network topology with at least one eigenvalue that lies in the instability region, the synchronized state is lost.} }
	\label{fig:circular}
\end{figure*}
%\end{widetext}

%\twocolumngrid

Before {{proceeding} in the quest {for}} the optimal network topological features that minimize the MSF, we {will} introduce two {simple} network {models}, shown in Fig.~\ref{fig:net_model}, {{for} which we can tune the directionality and the non-normality acting on a single parameter}. In the first case,  Fig.~\ref{fig:net_model} $a)$, we consider a bidirectional circulant network, i.e., a network whose adjacency matrix is circulant~\cite{circulant}, made by two types of links, one of weight $1$ forming a clockwise ring and the other {winding} a counterclockwise ring of tuneable weights $\epsilon$. The latter can vary in the interval $\varepsilon \in [0,1]$ exploring {in} this way the {possible topologies} from a fully symmetric case when $\varepsilon=1$ to a totally {mono-directed network} when $\varepsilon=0$. Since such a network is circulant, the adjacency matrix will be normal, a property that is inherited by the Laplace operator. On the contrary, if we remove two reciprocal links, respectively, of weights $1$ and $\varepsilon$, we obtain instead a non-normal network, as depicted in Fig.~\ref{fig:net_model} $b)$. In this case, the adjacency matrix is non-normal~\cite{trefethen}, a feature also reflected on the Laplacian matrix. Even in this case, we can tune the non-normality by varying the $\varepsilon$ parameter in the unitary interval as for the previous {case}, {this can be appreciated from the results} shown in Fig.~\ref{fig:net_model} $c)$ {where we report the normalized Henrici index, a well{-known} proxy of non-normality, as a function of $\varepsilon$}.
The main advantage of {using} the above network models is {the existence of} a basis of eigenvectors {for the Laplacian matrix}. In the first network model, this is due to the normality of {the} graph Laplacian, while in the second one {it is} because of the tridiagonal form of the coupling operator~\cite{footnote1}. This property is essential for the {applicability of the} MSF analysis, which is impossible otherwise.

\subsection{The case of normal directed networks}
\noindent
We start by considering the bidirected circular network and studying the linear stability of the synchronized state using the MSF analysis. The results shown in Fig.~\ref{fig:circular} $a)$, indicate that the network topology increasingly {contrasts} the stability {of the synchronous manifold} when the directionality increases. In fact, when the {MSF computed for the} directed network is compared to the symmetric case used as reference line (the continuous magenta curve), {we can always observe larger values, which moreover increase as $\varepsilon$ decreases (for the same fixed Laplacian eigenvalue)}. {Because of the circulant property of the Laplace matrix, its spectrum can be explicitly computed~\cite{malbor2}} $\Lambda_\alpha= 1+\varepsilon +(1+\varepsilon)\cos(2\alpha\pi/N) + i(1-\varepsilon)\sin(2\alpha\pi/N)$. One can easily notice that for $\varepsilon=0$, the spectrum distributes {uniformly} {onto the} unitary circle {centered} at $(1,0)$ as also shown in Fig. \ref{fig:circular} $b)$ in blue stars. On the other side, when $\varepsilon=1$, the network turns symmetric, making the spectrum real.

%\onecolumngrid

\begin{figure*}[t!]
	\centering
	\includegraphics[width=\textwidth]{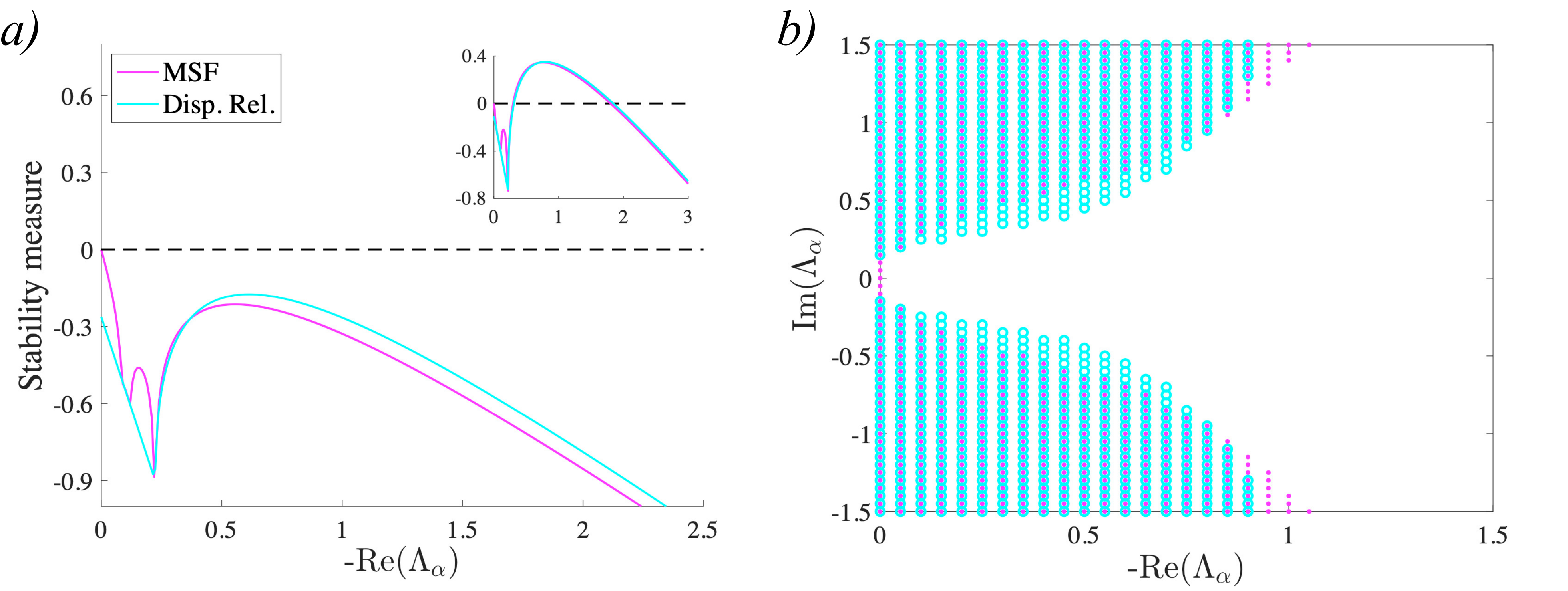}
	\caption{{$a)$} The {comparison} of the MSF and dispersion relation for the Brusselator with model parameters $b=3$, $c=1.8$, $D_{\varphi}=0.7$, $D_{\psi}=5$. We depict in {magenta} the MSF of the system in a limit cycle regime and {cyan} the dispersion relation of the {averaged} autonomous system. {Inset: Similar comparison for a set of parameters where the instability occurs. Notice also the lack of {continuity} of the stability {interval} of eigenvalues. $b)$ The same representation in the complex domain.} We see that for the chosen values of the parameters, the two approaches give an excellent {agreement} in predicting the instability {interval}.}
	\label{fig:confrontation}
\end{figure*}

%\twocolumngrid

{The MSF formalism ultimately relies on the maximum Lyapunov exponent, which despite having proved its validity in ruling out the chaotic behavior of dynamical system~\cite{strogatz}, remains {grounded on} numerical methods.} {To improve our analytical understanding of the problem}, we proceed by transforming {Eq.~\eqref{eq:MSF} into} an autonomous one, allowing {in} this way to deploy {the} spectral analysis {tools}. {This} method is part of the broader set of homogenization methods that aim at averaging a time-dependent {system {to obtain }a} time-independent one \cite{averaging}. Such methods have been found useful also for the stability analysis of synchronized states~\cite{rulk,chall}. The {resulting} autonomous version of the MSF is sometimes referred to as the dispersion relation~\cite{murray}. The  mathematical validity {of the proposed approximation} is grounded on the Magnus series expansion truncated at the first order \cite{chall}; hence, the set of {model} parameters {for which we expect a good agreement with the original model} corresponds to the case when higher-order terms are negligible. For more details, the interested reader should {consult}~\cite{chall}. In {formula}, it translates to
\begin{equation}
\boldsymbol{\mathcal{J}}(t)\longrightarrow \langle \boldsymbol{\mathcal{J}} \rangle_T =\frac{1}{T}\int_0^T \boldsymbol{\mathcal{J}}(\tau)d\tau\,.
\label{eq:average}
\end{equation}

Remarkably, as shown in Fig.~\ref{fig:confrontation}, this approximation yields qualitative results {in excellent agreement with the original model} for a specific range of parameters. An alternative to this approach is to apply a perturbative expansion near the bifurcation point, obtaining this way the time-independent Ginzburg-Landau normal form \cite{kuramoto}. However, the {effectivity of the} latter method is exclusively limited to parameters {values} very {close to} the stability threshold. In this sense, our approach is more general, both from {allowing a larger} set of parameters where the method remains valid, and at the same time, it is independent of the choice of the model compared to previous works \cite{dipatti}. The passage to an autonomous system is also essential in explaining the effect of the imaginary part of the Laplacian eigenvalues in the newly obtained stability function, the dispersion relation. It has been rigorously shown in \cite{malbor, malbor2} that the dispersion relation increases {proportional} to the magnitude of the imaginary part of the spectrum. We {already} observed similar results for the case of the MSF presented in Fig.~\ref{fig:circular}. We can {in} this way conclude that the averaging method sheds light on the {role of the directed topology} in the destabilization of a synchronized regime.}

\subsection{The case of non-normal directed networks}

\begin{figure*}[t!]
	\centering
	\includegraphics[width=\textwidth]{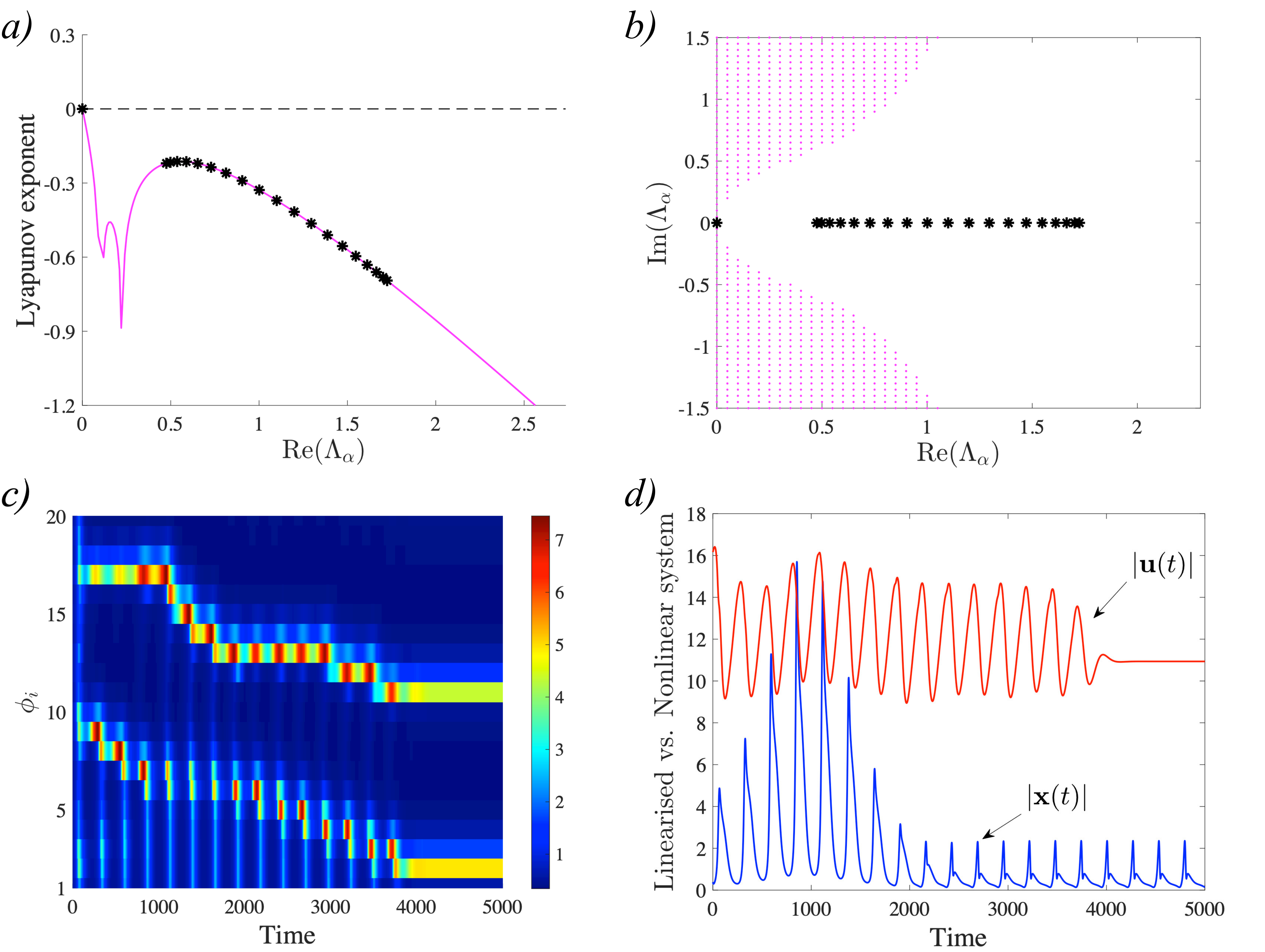}
	\caption{Desynchronization in a non-normal network. The parameters for the Brusselator model are as follows $b=2.5$, $c=1$, $D_{\varphi}=0.7$, $D_{\psi}=5$ on the (directed chain) non-normal network of $20$ nodes with $\varepsilon=0.1$ of Fig. \ref{fig:net_model} $b)$. As it can be observed from panels $a,)$ and $b)$, respectively, for the MSF and the stability region, the set of parameters is such that the MSF is neatly stable. Nevertheless, the instability occurs as shown by the pattern evolution in panel $c)$ at odd with the outcome that would have been expected from the symmetrized version. Such a result is strong evidence of the role of the network non-normality in the nonlinear dynamics of the system under investigation. {The mechanism that drives the instability in the non-normal linearized regime  manifests in the transition growth of the perturbations vector $\mathbf{x}(t)$ eq. \eqref{eq:compact}, the blue curve in panel $d)$, before the system relaxes to the oscillatory state of the equilibrium. Such growth might transform in a permanent instability for the nonlinear system $\textbf{u}(t)=\left[\boldsymbol{\varphi}(t),\boldsymbol{\psi}(t)\right]$, red curve.}}% Inset: The linearised response where now the asymptotic (residual) oscillatory behavior has been intentionally removed.}}
	\label{fig:NN}
\end{figure*}

\noindent
{The analysis performed in the previous section has been based on the study of the linearized system, in some cases, however,} such analysis is not sufficient to understand the outcome of the nonlinear system. In Fig.~\ref{fig:NN} {we consider again the MSF computed for the directed chain previously introduced (panel $b)$ of Fig.~\ref{fig:net_model}). 
{From Fig.~\ref{fig:NN} $b)$ one might naively} conclude that the system will synchronize, 
{since} the MSF {is non-positive} for all values of $\Re (\Lambda_\alpha)$. Moreover, the spectrum is completely real (see panel $b)$) and thus there cannot be any contribution from the imaginary part of the spectrum. However a direct inspection of the orbit behavior (panel $c)$) clearly shows that the system does not synchronize. Once the system is defined on a symmetric support, the synchronized behavior is recovered (panel $d)$).}
{This diversity of behavior is related to the non-normal property of the considered network, indeed it has been recently proved that such structural property can strongly alter the asymptotic behavior of networked systems~\cite{gao}. A finite perturbation about a stable equilibrium goes through a transient amplification {(see Fig. \ref{fig:NN} $d)$)} proportional to the level of non-normality before it is eventually reabsorbed in the linear approximation~\cite{trefethen}, while in the full non-linear system the finite perturbation could persist indefinitely. Up to now, this analysis has been limited to the case of autonomous systems; in this paper for the first time we extend it to the periodic time-dependent case making use of the homogenization process. This explains the permanent instability, shown in Fig.~\ref{fig:NN}, causing the loss of stability for the synchronized state.}

\begin{figure*}[t!]
	\centering
	\includegraphics[width=\textwidth]{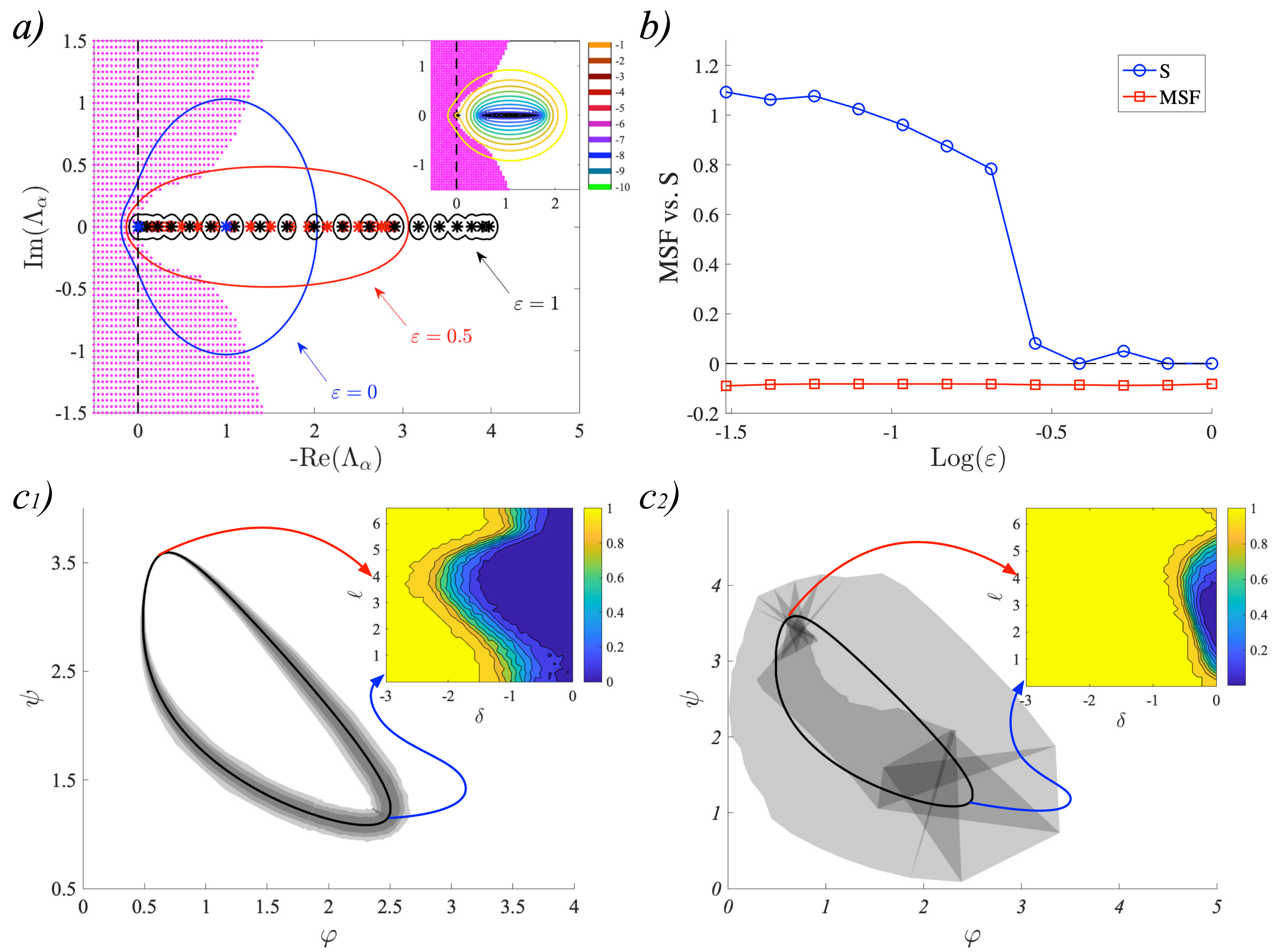}
	\caption{{$a)$} The pseudo-spectral description of the stability of the directed {chain of $20$ nodes} for the Brusselator model with $b=2.5$, $c=1.12$, $D_{\varphi}=0.7$, $D_{\psi}=5$, and an initial condition perturbation of the average magnitude $\delta=0.1$. We {show the pseudo-spectra for three different values of the control parameter $\varepsilon$} for the chain network, emphasizing the considerably large difference between the pseudo-spectra regions and the spectrum of the Laplacian matrix. {Inset: the pseudo-spectra for many other values of the perturbation magnitude $\delta$ for the chain with $\varepsilon=0.1$.} Notice {that although the eigenvalues do not lie inside the instability region due to the lack of an imaginary part, the pseudo-spectra might do.} {$b)$} The comparison between the expected outcome as predicted from the MSF and the actual outcome as measures by the standard deviation of the desynchronized pattern. {The stability basin (shaded grey) projected onto the limit cycle plane for the non-normal case, panel $c_1)$ and the symmetrized (normal) one, panel $c_2)$, calculated over $300$ different initial conditions (of the same averaged magnitude) and a perturbation whose maximum magnitude varies from $10^{-3}$ to $1$. Inset: In the $y$-axis we plot the points of limit cycle we perturb and in the $x$-axis the magnitude of the perturbation; the colormap gives the fraction of orbits that conserve the synchronized regime. It can be clearly noticed that the attraction basin for the non-normal network is strongly reduced, though not at the same amount compared to where the perturbation occurs.} }
	\label{fig:pseudo}
\end{figure*}

The non-normal dynamics study cannot be straightforwardly tackled with the analytical methods of the local stability, mostly because the instability occurs in a highly nonlinear {regime}. Such condition require a global analysis {that can be obtained using the numerical technique} based on a spectral perturbation {concept known as the} pseudo-spectrum. For a given matrix $\mathbf{A}$ {the latter} is defined as $\sigma(\mathbf{A}_\delta)= \sigma\left(\mathbf{A}+\mathbf{E}\right),$ {for all} $||\mathbf{E}||\leq\delta$ for where $\sigma(\cdot)$ represents the spectrum and $||\cdot||$ a given norm. {The package EigTool~\cite{eigtool} allows {us} to {compute and} draw in the complex plane the level curves of the pseudo-spectrum for a given value of $\varepsilon$.} Although the pseudo-spectrum is not sufficient to fully explain the system behavior, it is certainly of great utility in estimating the role of non-normality in the dynamics {outcomes}. In particular, in panel $b)$ {of Fig.~\ref{fig:pseudo} we report level curves} of the pseudo-spectrum for three different values of the {parameter $\varepsilon$ representing the} reciprocal links of the {directed} chain. {Notice} that by increasing the non-normality of the toy network, the pseudo-spectrum will also increase the chances {of} intersection with the instability region.  In panel $d)$ {of Fig.~\ref{fig:pseudo}}, we have shown a comparison between {a proxy of the presence of a synchronized state, i.e.} the standard deviation $S$~\cite{footnote2} of the {asymptotic orbit behavior} and the MSF demonstrating a clear different behavior{. For} all the considered values of $\varepsilon$ the MSF is always negative suggesting a stable synchronized state, on the other hand $S$ becomes positive and large for small enough $\varepsilon$, testifying a loss of synchronization. The {dependence on} the different values of the initial conditions is further shown in {panels $c_1)$ and $c_2)$}. As expected, the instability is more probable for both larger values of non-normality and magnitude of the initial conditions. {In particular, it can be observed that the synchronization basin of attraction is strongly reduced for the non-normal network compared to the normal one, and moreover its width varies along the limit cycle, implying that desynchronization will depend also on the point at which the perturbation starts.}

\section{Conclusions}
\noindent
In this paper, we have studied the quest for {the optimal conditions ensuring the stability of} synchronization dynamics in directed networks. {Such conditions determine} the design of a networked system that makes the synchronization regime as robust as possible. Previous results have proven that a strictly directed topology is necessary for the synchronized state's robustness. Based on the well-known Master Stability Function, it has been shown that directed tree-like networks are optimal for models with a {discontinuos interval of the Laplacian spectrum in the stability range of MSF}. Here, we have extended such results proving that they are generally independent of the {dynamic} model. Using an averaging procedure, we transformed the problem from a time-{dependent} (non-autonomous) to a time-invariant (autonomous) one. This method allows to prove that networks whose Laplacian matrix {exhibits a spectrum that} lacks an imaginary part are the most optimal. In general, the {loss of synchronization} {increases} with the magnitude of the imaginary part of the spectrum. {Secondly,} recent findings have shown that real-world networks present strong directed traits, resulting in a strong non-normality. This latter feature can play a very important role in the linear dynamics influencing the local stability of the synchronized state {through a} strong transient amplification of the perturbations. We have extended {the idea of non-normal dynamics to the case of } non-autonomous synchronization dynamics, revealing how network non-normality can drive the system to instability{, thus increasing the understanding of synchronization in complex networks. {We have also numerically quantified the effect of non-normality in driving the instability through the pseudo-spectrum technique.} }
In conclusion, we have analytically and numerically demostrated that there is no compelling recipe for optimal network architecture in order to conserve the synchronized state, but rather a trade-off between the network directedness and its non-normality. {We are aware that the interesting outcomes of the interaction of structrural non-normality networks with the fascinating synchronization phenomenon require deepper and further investigation (e.g. synchronization basin). In this sense, with this work we aim to initiate a new direction of research of the synchronization problem. }

\section*{Acknowledgements}
The authors would like to thank Duccio Fanelli for useful suggestions and feedbacks. R. M. is particularly grateful to Bram A. Siebert, Kleber A. Oliveira and Joseph D. O'Brien for interesting and helpful discussions and to the Erasmus+ program and the University of Limerick for funding his research visit in Limerick. The work of R.M. is supported by a FRIA-FNRS PhD fellowship, grant FC 33443, funded by the Walloon region. The {work  of} J.P.G. and M.A. is partly funded by Science Foundation Ireland (Grants No. 16/IA/4470, No. 16/RC/3918, No.  12/RC/2289 P2, and No. 18/CRT/6049) and cofunded under the European Regional Development Fund.


\begin{thebibliography}{4}
\bibitem{footnote1}
	This follows from the fact that every tridiagonal matrix is similar to a symmetric one and consequently. However, the set of eigenvectors is not the same; they are linearly independent in both cases.
\bibitem{footnote2}
	For a given species, e.g., $\varphi$, the standard deviation at a given time $t$ is $S(t)=\sqrt{\frac{1}{n-1}\sum_{i=1}^n\left(\varphi_i (t)-\hat{\varphi}(t)\right)}$, where $n$ is the number of nodes and $\hat{\varphi}(t)=\frac{1}{n}\sum_{i=1}^n \varphi_i(t)$. We can observe that such quantity is $0$ if the system is synchronized, i.e., $\varphi_i (t)=\hat{\varphi}(t)~\forall t$, while a value different from $0$ indicates that not every node has the same concentration at any given time.
\end{thebibliography}
\end{document}